\documentclass[epjc,twocolumn,nofootinbib,preprintnumbers,amsmath,amssymb]{revtex4-2}

\usepackage{graphicx}
\usepackage{dcolumn}
\usepackage{bm}
\usepackage[usenames]{color}
\usepackage{graphicx,epsfig}
\usepackage[colorlinks={true}]{hyperref}
\hypersetup{colorlinks=true,linkcolor=red,citecolor=blue,urlcolor=blue}
\usepackage{orcidlink}

\begin{document}

\title{
Comment on ``Environmental memory effects and quantum resource hierarchies in polarized hyperon--antihyperon systems''
}

\author{Saeed~Haddadi\orcidlink{0000-0002-1596-0763}}
\email{haddadi@ipm.ir}
\affiliation{
School of Particles and Accelerators,\\
Institute for Research in Fundamental Sciences (IPM),\\
P.O. Box 19395-5531, Tehran, Iran
}

\date{\today}

\begin{abstract}
We comment on the recent work [\href{https://doi.org/10.1140/epjc/s10052-026-16247-1}{Eur. Phys. J. C {\bf 86}, 988 (2026)}], which investigates quantum
resources in polarized hyperon--antihyperon systems using correlated
dephasing channels with memory. Although the use of experimentally reconstructed
spin-density matrices to characterize quantum correlations in hyperon
production may be a good motivation, we argue that the interpretation of the
hyperon--antihyperon pair as an open quantum system undergoing correlated
environmental decoherence is not physically established. In particular, no
physical environment or system--environment interaction responsible for the
assumed correlated dephasing channel is identified, and hadronization cannot
simply be interpreted as such an environmental dephasing process. Consequently,
the claimed non-Markovianity, information backflow, and memory-assisted
protection of quantum resources should be regarded as properties of a
phenomenological channel rather than established physical effects in
hyperon--antihyperon production. We also identify several technical and conceptual issues concerning the dephasing dynamics, entanglement quantification, basis dependence of coherence, and the claimed hierarchy of quantum resources. These issues call for a substantial revision of the
physical interpretation and quantitative conclusions of the work. We hope
that the present Comment will help clarify the distinction between
phenomenological quantum-channel modeling and physically established
dynamical mechanisms in high-energy particle systems, and thereby provide
useful guidance for future studies and help avoid similar conceptual and
technical issues.
\end{abstract}

\maketitle

%%%%%%%%%%%%%%%%%%%%%%%%%%%%%%%%%%%%%%%%%%%%%%%%%%%%%%%%%%%%%%%%%%%%%%%

The recent paper by Bachain, Amazioug, and Ahl Laamara~\cite{Bachain} investigates logarithmic negativity, geometric quantum discord, and $l_1$-norm coherence in polarized hyperon--antihyperon pairs produced through
\begin{equation}
e^+e^-\rightarrow J/\psi\rightarrow Y\bar Y,
\end{equation}
with $Y=\Lambda,\Sigma^+,\Xi^-,\Xi^0$. The authors combine experimentally determined production parameters with a correlated dephasing model containing a memory parameter and interpret the resulting dynamics in terms of Markovian and non-Markovian evolution, information backflow, and memory-assisted preservation of quantum resources.

The study addresses an intersection between particle physics and quantum information science. In particular, the use of experimentally reconstructed spin density matrices to characterize entanglement and other quantum information quantities in high-energy reactions is motivated. However, we believe that several assumptions underlying the dynamical interpretation of Ref.~\cite{Bachain} require substantial clarification. Our principal concern is more fundamental than the particular choice of quantum resource measure: it concerns whether an open quantum system description with a correlated environmental dephasing channel can physically be assigned to the produced hyperon--antihyperon pair in the process under consideration.

This issue is especially important because essentially all of the dynamical conclusions of Ref.~\cite{Bachain}---including the claimed information backflow, recurrent revivals, environmental-memory protection, and the resulting hierarchy of resource robustness---follow from the assumed correlated dephasing map. Moreover, we point out several technical and
conceptual problems, including an incorrect implementation of the colored
dephasing function, the misidentification of negativity as logarithmic
negativity and an associated factor-of-two inconsistency, the basis
dependence of the $l_1$-norm of coherence under the employed local-unitary
transformations, and an unsupported interpretation of the numerical ordering
of coherence, entanglement, and geometric discord as a universal hierarchy
of quantum resources.
\\
\\
\textbf{Remark~1.} A spin-$1/2$ hyperon and antihyperon can certainly be represented by two two-dimensional spin Hilbert spaces, and their joint state can consequently be represented by a $4\times4$ density matrix. This mathematical representation is useful and provides a natural language for describing experimentally measured polarization and spin correlation observables.
The important question, however, is not whether a $4\times4$ density matrix can be written, but whether its subsequent evolution can be physically interpreted as the reduced dynamics of an open quantum system.
In the reaction
$e^+e^-\rightarrow J/\psi\rightarrow Y\bar Y$,
the hyperon and antihyperon are produced during a single relativistic scattering and decay process. Once produced, they propagate as unstable relativistic particles and subsequently decay through weak interactions. The spin density matrix reconstructed from angular distributions is primarily a characterization of the production process and its polarization and spin correlation structure. This is fundamentally different from the usual open system scenario in which a quantum subsystem is initially prepared and then evolves through a specified interaction with an external environment.

To interpret an open system (S), one usually needs at least a physical description of the environment (E), the system-environment (SE) interaction, and a corresponding reduced dynamical map~\cite{Breuer},
\begin{equation}
\rho_\mathrm{S}(t)=\operatorname{Tr}_\mathrm{E}
\left[
U_{\mathrm{SE}}(t)\rho_{\mathrm{SE}}(0)U_{\mathrm{SE}}^{\dagger}(t)
\right].
\end{equation}
Equivalently, in a phenomenological description one needs a physically motivated map
$
\rho_\mathrm{S}(t)=\Phi_t[\rho_\mathrm{S}(0)]
$
whose parameters can be related to properties of an actual environment.

No such physical construction is provided in Ref.~\cite{Bachain}.
The authors instead introduce a \textit{correlated dephasing channel}~\cite{Ref4,HuZhou2019,Daffer2004} and subsequently interpret its parameters as environmental effects in the hyperon system. In particular, the memory parameter $\mu$ is treated as a measure of environmental correlations, while the parameter $\tau$ controls the Markovian or non-Markovian behavior. However, the physical origin of these parameters in $J/\psi\rightarrow Y\bar Y$ production is not established.
This distinction is essential. A phenomenological quantum channel can be applied \textit{mathematically} to an experimentally reconstructed density matrix. Nevertheless, this mathematical construction does not imply that the assumed channel represents a physical dynamical process governing the hyperon--antihyperon system during its propagation and decay.
\\
\\
\textbf{Remark~2.}
The channel adopted in Ref.~\cite{Bachain} is a correlated dephasing channel with a memory parameter. Such channels have a clear interpretation in quantum information theory~\cite{Ref4}: two systems interact with a common or correlated environment, and environmental correlations modify the resulting reduced dynamics.
For a hyperon--antihyperon pair, however, what is the corresponding common environment?
The paper does not identify one. Several possible physical effects might be present in a realistic collider experiment: hadronization, electromagnetic interactions, residual strong interactions, interactions with detector material, background processes, finite detector resolution, acceptance effects, and decay dynamics. But these effects \textit{cannot} simply be equated with an externally imposed correlated dephasing reservoir.

Notice that hadronisation is not naturally an environmental channel acting on an already prepared hyperon--antihyperon qubit pair. Hadronization is part of the nonperturbative QCD dynamics through which the observed hadronic state is produced. It therefore belongs to the production mechanism itself rather than constituting a subsequent reservoir acting on a pre-existing two-qubit system.
Likewise, detector interactions and detector resolution affect the measurement process. They can produce experimental smearing, acceptance effects, reconstruction biases, and statistical uncertainties, but this does not automatically imply a physical quantum dynamical map acting on the hyperon--antihyperon spin state.
Consequently, the central distinction should be made between \textit{physical production and decay dynamics} and \textit{a phenomenological quantum channel} applied to the reconstructed state. Unfortunately, Ref.~\cite{Bachain} effectively treats these two notions as equivalent.
\\
\\
\textbf{Remark~3.}
The above issue becomes particularly serious in the discussion of non-Markovianity.
The paper interprets the oscillatory behavior of the decoherence function as environmental memory and repeatedly describes the associated revivals as information backflow from the environment to the hyperon pair. In the theory of open quantum systems, however, non-Markovianity is a property of a dynamical map. It can be characterized, for example, through completely positive (CP) divisibility or information-backflow measures such as the Breuer--Laine--Piilo criterion~\cite{Breuer,BreuerLainePiilo,Rivas}. The physical interpretation requires a system whose reduced dynamics is generated by coupling to environmental degrees of freedom.
In the present problem, no experimentally identifiable environment is shown to generate the assumed map. Therefore, the oscillations generated by the chosen function $K(t)$ establish only that the mathematical channel itself produces oscillatory dynamics. They do not establish that the physical hyperon system experiences environmental information backflow.
This point was also emphasized in our previous Comment~\cite{HaddadiPRD}, where we argued that, for unstable relativistic hyperons, the physical origin and operational meaning of environmental memory must be established before assigning non-Markovianity to the reconstructed spin state.
\\
\\
\textbf{Remark~4.}
Hyperons are unstable particles. Their evolution is governed by their production dynamics, relativistic propagation, and intrinsic weak-interaction decay. This is qualitatively different from the evolution of a stable qubit coupled to a stationary external reservoir.
The authors~\cite{Bachain} introduce dimensionless time variables and investigate, for example, $\tau=5$ and $\tau=0.2$. However, no physical mapping is provided between the model time and the proper time or laboratory propagation time of the hyperon.

For an experimentally meaningful dynamical claim, one should establish a relation of the form
\begin{equation}
t_{\rm model}\longleftrightarrow t_{\rm proper}
\longleftrightarrow t_{\rm lab},
\end{equation}
and demonstrate that the predicted revivals occur within the physically accessible lifetime and decay region of the relevant hyperon.
Without such a mapping, the statement that non-Markovian memory protects quantum resources in the experimentally produced hyperon system remains a property of the chosen mathematical model rather than an experimentally testable prediction.
\\
\\
\textbf{Remark~5.}
There is a significant mathematical inconsistency in the implementation of the
colored dephasing dynamics adopted in Ref.~\cite{Bachain}. The dynamical function used by the authors in Eqs.~(35) and (36) is not the correct form of the colored dephasing model on which their analysis is based.

In the standard random telegraph noise model employed for this type of
dephasing channel, the decoherence function in the non-Markovian regime
$\tau>1/4$ is given by~\cite{HuZhou2019}
\begin{equation}
\Phi(t)=
e^{-t/(2\tau)}
\left[
\cos\left(\frac{ut}{2\tau}\right)
+
\frac{1}{u}
\sin\left(\frac{ut}{2\tau}\right)
\right],
\end{equation}
where
\begin{equation}
u=\sqrt{\left|1-16\tau^2\right|}.
\end{equation}
For the Markovian regime $\tau<1/4$, the corresponding expression is obtained
by replacing the trigonometric functions by the appropriate hyperbolic
functions. This is the form given, for example, in Refs.~\cite{HuZhou2019,Daffer2004}.
In contrast, the authors~\cite{Bachain} introduces
\begin{equation}
u=\frac{1}{2\tau},
\qquad
v=\sqrt{1-u^2},
\end{equation}
and subsequently employs $u$ and $v$ in Eqs.~(35) and (36). This is not the
correct parametrization of the colored dephasing function described above.
In particular, the quantity controlling the transition between the
trigonometric and hyperbolic forms is
$1-16\tau^2$,
rather than $1-u^2$ with $u=1/(2\tau)$.

Indeed, the correct crossover between the oscillatory and non-oscillatory
regimes follows from
\begin{equation}
1-16\tau^2=0,
\end{equation}
which gives
\begin{equation}
\tau=\frac14.
\end{equation}
Thus, the separation between the non-Markovian and Markovian regimes,
$\tau>1/4$ and $\tau<1/4$, respectively, is consistent with the standard
colored dephasing model. The difficulty in Ref.~\cite{Bachain} is therefore not the
value of the crossover itself, but the incorrect functional form and
parametrization used in Eqs.~(35) and (36).

This error is consequential because the decoherence function $\Phi(t)$ ($K(t)$ in Ref.~\cite{Bachain}) enters
directly into the time-dependent density matrix and, consequently, into all
of the quantum-resource measures investigated in the paper. In the correlated
dephasing model, for example, the output density matrix depends on
$\Phi(t)$ through both the single-qubit coherences and the correlated term
\begin{equation}
\omega(t,\mu)=\Phi^2(t)+[1-\Phi^2(t)]\mu,
\end{equation}
where $\omega(t,\mu)=\eta(t)$ as seen in Eq.~(38). Therefore, an incorrect expression for $\Phi(t)$ propagates directly into
the predicted time dependence of coherence, discord, and entanglement.

Moreover, the distinction between the two regimes has a definite physical
origin in the underlying random telegraph noise model. The damped oscillatory
behavior in the non-Markovian regime originates from the trigonometric form of
$\Phi(t)$, whereas the Markovian regime is characterized by the corresponding
hyperbolic functions and a monotonic decay. The reference model explicitly
identifies $\tau>1/4$ as the non-Markovian regime and $\tau<1/4$ as the
Markovian regime~\cite{HuZhou2019}. Thus, the expressions used in Ref.~\cite{Bachain}
should be rederived from the underlying dephasing model rather than obtained
by the substitutions $u=1/(2\tau)$ and $v=\sqrt{1-u^2}$.

We therefore believe that Eqs.~(35) and (36) of Ref.~\cite{Bachain} should be corrected
and that all subsequent numerical results based on these equations should be
recalculated. In particular, the time-dependent curves presented for the
quantum-resource measures may change quantitatively, and potentially
qualitatively, once the correct decoherence function is used.
\\
\\
\textbf{Remark~6.}
A separate technical problem concerns the definition of the principal entanglement measure.
The paper~\cite{Bachain} states that it uses logarithmic negativity but defines
\begin{equation}
E_N(\rho)
=
\frac{\|\rho^{T_B}\|_1-1}{2}.
\end{equation}
This is the standard definition of the negativity~\cite{Vidal2002},
not logarithmic negativity.
The logarithmic negativity is instead defined as
\begin{equation}
E_{\mathcal N}(\rho)
=
\log_2\|\rho^{T_B}\|_1,
\end{equation}
up to the convention for the logarithm.
This distinction is important because the two measures have different numerical values and different normalization properties. As a result, all references to ``logarithmic negativity'' throughout the paper should be corrected unless the authors actually intend to use the ordinary negativity.

There is additionally an internal factor-of-two inconsistency. The paper first writes
\begin{equation}
E_N(\rho^X)
=
\frac{\|\rho^{T_B}\|_1-1}{2}
=
-\sum_i\mu_i,
\end{equation}
and then obtains
\begin{equation}
E_N(\rho^X)
=
\max\{0,-2\mu_{\min}\}.
\end{equation}
This is consistent for a two-qubit state with at most one negative eigenvalue.
However, after defining
\begin{equation}
\mu_{\min}=\min\{e_1,e_2,e_3,e_4\},
\end{equation}
the paper states
\begin{equation}
E_N(\rho^X)
=
\max[0,-\min\{e_1,e_2,e_3,e_4\}],
\end{equation}
which is missing the factor of ``2''.
Thus, Eq.~(46) in Ref.~\cite{Bachain} does not follow from Eqs.~(42) and (43).
This should be corrected before any quantitative comparison involving the entanglement measure is considered reliable.
\\
\\
\textbf{Remark~7.}
The paper repeatedly claims a hierarchy among coherence, entanglement, and
geometric quantum discord, concluding that entanglement is the most robust
quantum resource, followed by coherence and then geometric discord (e.g. $\mathcal{L}_N>C_{l_1}>\mathcal{D}_G$ in Fig.~29), and $C_{l_1}>\mathcal{L}_N>\mathcal{D}_G$ in Fig.~30.
However, this conclusion is not supported by the hierarchy of quantum correlations
in quantum resource theory, in which we have: quantum coherece $\supseteq$ quantum discord $\supseteq$ entanglement~\cite{Wiseman2007}.
Moreover, the three quantities are fundamentally different quantum-resource
measures, with different operational meanings, normalizations, and
invariance properties. Therefore, their numerical magnitudes cannot in
general be interpreted as defining a universal hierarchy of quantum
resources. Contrary to what the authors claim throughout the article~\cite{Bachain}, the reported ordering is not preserved across their numerical results; rather, it changes with the chosen parameters and time.
This is particularly important because the authors interpret the numerical
ordering of the three measures as evidence for a hierarchy of robustness.
A change in the ordering with the model parameters or time demonstrates
that such a hierarchy cannot be regarded as an intrinsic property of the
quantum resources themselves.
\\
\\
\textbf{Remark~8.}
A more fundamental problem concerns the calculation of coherence itself.
The authors first perform local-unitary transformations to diagonalize the correlation tensor and obtain the canonical form
\begin{equation}
\Theta'
=
\begin{pmatrix}
1&0&0&B_z\\
0&\lambda_1&0&0\\
0&0&\lambda_2&0\\
B_z&0&0&\lambda_3
\end{pmatrix},
\end{equation}
and the corresponding $X$ state~\cite{Bachain}.
They observe that local unitary transformations preserve entanglement. However, this argument cannot be extended to the $l_1$-norm coherence.
The $l_1$-norm coherence
\begin{equation}
C_{l_1}(\rho)=\sum_{i\neq j}|\rho_{ij}|
\end{equation}
is explicitly basis dependent~\cite{Baumgratz2014}. It is not invariant under arbitrary local unitary transformations.
Therefore, the coherence calculated from the canonical $X$ state is generally not equal to the coherence of the original experimentally defined spin density matrix.

This point is crucial. The local rotations used to diagonalize $C$ may change the off-diagonal matrix elements and hence change $C_{l_1}$, even though they leave entanglement invariant.
Consequently, Eq.~(57) in Ref.~\cite{Bachain},
\begin{equation}
C_{l_1}(\rho^X)
=
\frac12
\left(
|\lambda_1-\lambda_2|
+
|\lambda_1+\lambda_2|
\right),
\end{equation}
does not by itself provide the $l_1$-norm coherence of the physical hyperon state in the experimentally specified spin basis.
The authors should either calculate the coherence in a physically specified basis or explicitly state that they are evaluating a basis-dependent coherence in the canonical local reference frame. The two interpretations should not be conflated.
This issue directly affects the central conclusion concerning the robustness of coherence.
\\
\\
\textbf{Remark~9.}
Another conceptual distinction is required between coherence and interparticle correlations.
The quantity
$
C_{l_1}(\rho_{Y\bar Y})
$
measures coherence of the joint density matrix in a selected product basis. It does not, by itself, isolate the genuinely correlated component of coherence between $Y$ and $\bar Y$.
The total coherence may contain contributions associated with local coherence as well as correlated coherence. Therefore, comparing the total $l_1$-norm of coherence directly with bipartite entanglement or discord and interpreting the larger numerical value as evidence that ``coherence is more robust'' is not conceptually straightforward.
If the physical objective is to quantify quantum correlations between the hyperon and antihyperon, a correlated-coherence measure or another explicitly bipartite quantity would provide a more appropriate comparison.
\\
\\
\textbf{Remark~10.}
The authors use experimentally measured BESIII values of $\alpha_\psi$ and $\Delta\Phi$ and consequently argue that the predicted effects are experimentally compatible with BESIII and future facilities (see Table~1 in Ref.~\cite{Bachain}).
However, the use of experimental input parameters in the initial density matrix does not establish the experimental accessibility of the subsequent open system dynamics.
In particular, the parameters governing the environmental channel, including $\tau$ and $\mu$, are not extracted from experimental data. They are model parameters.
The distinction is therefore
\textit{experimentally measured production parameters}
versus
\textit{assumed environmental parameters}.
The former are physical inputs; the latter require an independent physical interpretation and experimental determination before the predicted memory effects can be regarded as experimentally testable properties of the hyperon system.
Furthermore, the uncertainties in the BESIII input parameters should be propagated through the nonlinear functions defining the various quantum-resource measures. The fact that central experimental values are used in the density matrix is not sufficient to demonstrate that the predicted effects are resolvable within experimental uncertainties.
A Monte Carlo propagation of the uncertainties in $\alpha_\psi$ and $\Delta\Phi$ into the density matrix and the resulting quantum-resource measures would substantially strengthen such an experimental claim (see~\cite{haddadiprb2026}).
\\
\\
\textbf{Remark~11.}
The experimentally reconstructed hyperon density matrix is obtained from decay distributions and spin observables. This provides access to the spin correlations of the produced pair.
However, the paper subsequently introduces
$
\rho_{Y\bar Y}(t)
$
and interprets its time dependence as environmental evolution.
An experimentally meaningful observation of the predicted revivals would require a concrete prescription connecting measured quantities to
$
\rho_{Y\bar Y}(t)
\rightarrow
C_{l_1}(t),\,
D_G(t),\,
E_{\mathcal N}(t).
$
In particular, the paper should explain how the model time variable is reconstructed experimentally, how the hyperon proper time is incorporated, how the decay distributions are modified by the proposed channel, and how the environmental parameters would be extracted from data.
Without such a prescription, the figures demonstrate the behavior of a mathematically evolved density matrix rather than a demonstrated time-dependent observable of the collider process.
\\
\\
\textbf{Remark~12.}
The paper identifies $\mu=0$ with an uncorrelated environment and $\mu=1$ with a fully correlated environment, and shows that increasing $\mu$ protects the quantum resources~\cite{Bachain}.
Within quantum-channel theory, this construction is meaningful. However, the physical interpretation of $\mu$ for a hyperon--antihyperon pair remains unclear.
In particular, it should be specified whether $\mu$ represents temporal correlations between successive interactions, spatial correlations of a common environment acting on the hyperon and antihyperon, correlations between two applications of a quantum channel, or some phenomenological effective parameter encoding experimental disturbances.
These possibilities are physically distinct.
For the reaction considered in Ref.~\cite{Bachain}, no mechanism is identified that would generate a common reservoir with a tunable correlation parameter $\mu$. Consequently, the observed enhancement at $\mu\rightarrow1$ should not be interpreted as evidence that the physical hyperon system possesses such environmental correlations.
\\
\\
\textbf{Remark~13.}
In Ref.~\cite{HaddadiPRD}, we pointed out that the produced $\Lambda\bar\Lambda$ pair originates from a single scattering event and subsequently propagates as unstable relativistic particles, while the experimentally reconstructed density matrix is primarily determined by production amplitudes and decay kinematics. We argued that the introduction of correlated quantum channels and non-Markovianity requires an additional physical mechanism and should not be inferred solely from the existence of spin correlations.
The work~\cite{Bachain} adopts essentially the same conceptual framework and, unfortunately, extends it to several hyperon channels. The extension does not resolve the underlying physical problem. The broader scope makes the issue more significant because the paper now interprets the phenomenological channel as a mechanism for protecting quantum resources across several experimentally relevant production processes.
\\
\\
\textbf{Conclusion.}
The study of quantum correlations in hyperon--antihyperon production is a promising direction at the interface of particle physics and quantum information. The spin-density-matrix formalism used to characterize the polarization and correlations generated in
$
J/\psi\rightarrow Y\bar Y
$
is well motivated, and experimentally measured spin observables provide a valuable basis for investigating entanglement and related quantities.

Our concern is specifically with the subsequent interpretation of this static production state as an open bipartite quantum system undergoing correlated environmental decoherence.
The principal difficulty is that Ref.~\cite{Bachain} does not identify a physical environment, a system--environment interaction, or an experimentally established dynamical map that would generate the assumed correlated dephasing channel. Consequently, the memory parameter $\mu$, the non-Markovian parameter $\tau$, and the associated information-backflow interpretation do not currently have a demonstrated physical correspondence in the hyperon--antihyperon production process.
This issue is particularly important because the principal conclusions of Ref.~\cite{Bachain} concerning memory-assisted protection, recurrent revivals, and the hierarchy of quantum-resource robustness are direct consequences of the assumed open-system dynamics.

In addition to this fundamental issue, we have identified several technical and conceptual problems: the quantity defined in Eq.~(40) is negativity rather than logarithmic negativity; Eq.~(46) contains a factor-of-two inconsistency; the stated Markovian/non-Markovian boundary is inconsistent with the definitions of $u$ and $v$; the $l_1$-norm of coherence is basis dependent and therefore cannot be assumed invariant under the local-unitary transformations used to obtain the canonical $X$ state; and the numerical ordering of different quantum measures should not be interpreted as a universal resource-theoretic hierarchy.

We therefore believe that the manuscript's conclusions should be substantially qualified. Quantum information measures can certainly be used to characterize experimentally reconstructed spin density matrices in high-energy processes. However, a distinction must be maintained between a mathematical probe of how a density matrix responds to a phenomenological quantum channel and a physically established open system dynamics of the underlying particle system.
For the hyperon--antihyperon system, the latter requires an explicit physical model of the relevant environmental degrees of freedom and their interaction with the hyperon spins, together with a connection between the resulting dynamical map and experimentally measurable observables. In the absence of such a construction, the effects presented in Ref.~\cite{Bachain} should be regarded as consequences of the assumed phenomenological correlated dephasing model rather than as established environmental memory effects in polarized hyperon production.
\\
\\
\\
\textbf{Conflict of interest:}
There are no known competing financial interests.
\\
\textbf{Data Availability Statement:}
This manuscript has no associated data.
\\
\textbf{Code Availability Statement:}
The manuscript has no associated code/software.
\\
\textbf{Funding:}
No funding.

%%%%%%%%%%%%%%%%%%%%%%%%%%%%%%%%%%%%%%%%%%%%%%%%%%%%%%%%%%%%%%%%%%%%%%%

\end{document}